\documentclass{appolb}
\usepackage{epsfig}
\begin{document}

\title{ Isospin asymmetry in nuclei and nuclear symmetry energy }
{
\author{ Tapan Mukhopadhyay \thanks{E-mail: tkm@veccal.ernet.in} and D.N. Basu \thanks{E-mail: dnb@veccal.ernet.in}  
\address{Variable Energy Cyclotron Centre, 1/AF Bidhan Nagar, Kolkata 700 064, India\\}}
\vskip 0.2cm
\maketitle
\begin{abstract}

      The volume and surface symmetry parts of the nuclear symmetry energy and other coefficients of the liquid droplet model are determined from the measured atomic masses by the maximum likelihood estimator. The volume symmetry energy coefficient extracted from finite nuclei provides a constraint on the nuclear symmetry energy. This approach also yields the neutron skin of a finite nucleus through its relationship with the volume and surface symmetry terms and the Coulomb energy coefficient. The description of nuclear matter from the isoscalar and isovector components of the density dependent M3Y effective interaction provide a value of the symmetry energy that is consistent with the empirical value of the symmetry energy extracted from measured atomic masses and with other modern theoretical descriptions of nuclear matter.
 
\end{abstract}


\section{Introduction}
\label{section1}

      The investigation of constraints of nuclear symmetry energy has recently received new impetus with the plans to construct a  new accelerator facility (FAIR) at GSI Dramstadt. The nuclear symmetry energy (NSE) plays a central role in determining to a large extent the equation of state (EoS) of isospin asymmetric nuclear matter, proton fraction in neutron stars and neutron skin in heavy nuclei and it enters as an input to the heavy ion reactions \cite{Li98},\cite{Li02}. Various many body calculations using a realistic nucleon-nucleon (NN) interaction as input (Brueckner-Hartree-Fock \cite{Ba01} or Dirac-Brueckner-Hartree-Fock \cite{Le98} and the variational method \cite{Ak98} etc.) lead to rather different results for the symmetry energy. In view of the large differences present between various calculations of the symmetry energy even at subsaturation densities, the question arises naturally whether one can obtain empirical constraints from finite nuclei. As the degree of isospin diffusion in heavy-ion collisions at intermediate energies is affected by the stiffness of the nuclear symmetry energy, these reactions provide constraints on the behaviour of the nuclear symmetry energy at subsaturation densities \cite{Ch05}. Traditionally, the symmetry energy of nuclei at saturation density is extracted by fitting ground state masses with various versions of liquid drop mass formula (LDM). To this end one needs to decompose the symmetry term of LDM into the bulk (volume) and surface terms \cite{Da03} along the lines of the liquid droplet model and identify the volume symmetry energy coefficient as the symmetry energy derived from infinite nuclear matter at saturation density. 
The coefficients of liquid droplet model are extracted by employing the maximum likelihood estimator method.   

      In the present work, the nuclear symmetry energy is calculated theoretically using the isoscalar and isovector components of M3Y-Reid-Elliott effective interaction supplemented by a zero range pseudo-potential along with the density dependence (DDM3Y) and its value at saturation density is compared with the volume symmetry energy coefficient extracted from a fit to the atomic mass excesses from the latest mass table \cite{Au03}. The M3Y interaction was derived \cite{Be77} by fitting its matrix elements in an oscillator basis to those elements of the G-matrix obtained with the Reid-Elliott soft-core NN interaction. The ranges of the M3Y forces were chosen to ensure a long-range tail of the one-pion exchange potential as well as a short range repulsive part simulating the exchange of heavier mesons \cite{Sa79}. The zero-range pseudo-potential represented the single-nucleon exchange term while the density dependence accounted for the higher order exchange effects and the Pauli blocking effects. The real part of the proton-nucleus interaction potential obtained by folding in the density distribution function of the interacting nucleus with the DDM3Y effective interaction is found to provide good descriptions of elastic and inelastic scatterings of high energy protons \cite{Gu05} and proton radioactivity \cite{BCS05}. In the present work, the results obtained for the symmetry energy from DDM3Y interaction are consistent with the results obtained by the fitting of masses to the liquid droplet model.

\section{The nuclear symmetry energy}
\label{section2}

      The nuclear EoS can be expanded in terms of isospin asymmetry X as

\begin{equation}
 \epsilon(\rho,X) = \epsilon(\rho,0) + S(\rho) X^2 + S_1(\rho) X^4 + S_2(\rho) X^6 + .~.~.~.~.
\label{seqn1}
\end{equation}
\noindent
which provides the standard definition of the nuclear symmetry energy $S(\rho)$ where $\epsilon(\rho,X)$ is the energy per nucleon of nuclear matter with isospin asymmetry $X = \frac{\rho_n - \rho_p} {\rho_n + \rho_p},~~~~\rho = \rho_n+\rho_p,$ where $\rho_n$, $\rho_p$ and $\rho$ are the neutron, proton and nucleonic densities respectively. The DDM3Y effective NN interaction is given by 

\begin{equation}
 v_{00}(s,\rho, \epsilon) = t_{00}^{M3Y}(s, \epsilon) g(\rho),~~~~v_{01}(s,\rho, \epsilon) = t_{01}^{M3Y}(s, \epsilon) g(\rho)
\label{seqn2}
\end{equation}   
\noindent
where $s$ is distance between two interacting nucleons, $g(\rho) = C (1 - \beta\rho^{2/3})$ is the density dependence and the isoscalar $t_{00}^{M3Y}$ and the isovector $t_{01}^{M3Y}$ components of M3Y interaction potentials \cite{Sa79}, \cite{BCS05} supplemented by zero range potentials are given by 

\begin{equation}
 t_{00}^{M3Y}(s, \epsilon) = 7999\frac{\exp( - 4s)}{4s} - 2134\frac{\exp( - 2.5s)}{2.5s} - 276 (1 - \alpha\epsilon)\delta(s)
\label{seqn3}
\end{equation}   
\noindent
and

\begin{equation}
  t_{01}^{M3Y}(s, \epsilon) =  -4886\frac{\exp( - 4s)}{4s} + 1176\frac{\exp( - 2.5s)}{2.5s} + 228 (1 - \alpha\epsilon)\delta(s)
\label{seqn4}
\end{equation}   
\noindent
respectively, where the energy dependence parameter $\alpha$=0.005/MeV. Based on the Hartree or mean field assumption and using the DDM3Y interaction, the EoS can be derived as 

\begin{equation}
 \epsilon(\rho,X) = [\frac{3\hbar^2k_F^2}{10m}] F(X) + (\frac{\rho J_v C}{2}) (1 - \beta\rho^{2/3}),
\label{seqn5}
\end{equation}
\noindent
where $k_F^3 = 1.5\pi^2\rho$, $J_v=J_{v00} + X^2 J_{v01}$, $J_{v00}(\epsilon) = \int \int \int t_{00}^{M3Y}(s, \epsilon) d^3s$ and $J_{v01}(\epsilon) = \int \int \int t_{01}^{M3Y}(s, \epsilon) d^3s$ represent the volume integrals of the isoscalar and isovector parts of the M3Y interaction and
 
\begin{equation}
 F(X) = [\frac{(1+X)^{5/3} + (1-X)^{5/3}}{2}],
\label{seqn6}
\end{equation}   
\noindent
where $m$ is the nucleonic mass equal to 938.91897 $MeV/c^2$. In nuclear matter, $\epsilon(\rho,X) = \epsilon(\rho,0) + \epsilon^\prime(\rho,X)$ so that $\epsilon^\prime(\rho,X) = \epsilon(\rho,X) - \epsilon(\rho,0) \approx S(\rho) X^2$ for other terms being small. Therefore, an alternative physical definition of the nuclear symmetry energy \cite{Kl06} is the energy required per nucleon to change the symmetric nuclear matter (SNM) to pure neutron matter (PNM) which is given by

\begin{equation}
 S(\rho)=\epsilon_{PNM} -\epsilon_{SNM}=(2^{2/3} - 1)\frac{3}{5}E^0_F(\frac{\rho}{\rho_0})^{2/3}+\frac{C}{2} \rho (1 - \beta\rho^{2/3}) J_{v01}
\label{seqn7}
\end{equation}
\noindent
where $\epsilon_{PNM}=\epsilon(\rho,1)$, $\epsilon_{SNM}=\epsilon(\rho,0)$ are the energy per particle for PNM and SNM respectively, $\rho_0$ is the saturation nucleonic density, $E^0_F=\frac{\hbar^2k_{F_0}^2}{2m}$ is the Fermi energy for the SNM in the ground state with $k_{F_0}$ as the corresponding Fermi momentum. The constants of density dependence $C$ and $\beta$ of the effective interaction are obtained by reproducing the saturation energy per nucleon and the saturation density of SNM \cite{Ba04}. In ref.\cite{Ba04} the symmetric nuclear matter properties such as its EoS and incompressibility were calculated whereas in this work we have extended that for asymmetric nuclear matter in order to calculate the nuclear symmetry energy. 

      The first term of the right hand side of Eq.(7) is the kinetic energy contribution whereas the second term is the potential energy contribution and accounts for the nuclear interaction. If one uses definition of Eq.(1) for the nuclear symmetry energy, then the second term remains unaltered while the the first term reduces by about five percent to $\frac{5}{3^2}[\frac{3}{5}E^0_F(\frac{\rho}{\rho_0})^{2/3}]$. The standard definition of Eq.(1) also provides $S_1(\rho) = \frac{5}{3^5}[\frac{3}{5}E^0_F(\frac{\rho}{\rho_0})^{2/3}]$ and $S_2(\rho) = \frac{35}{3^8}[\frac{3}{5}E^0_F(\frac{\rho}{\rho_0})^{2/3}]$ etc. Interestingly, the definition of nuclear symmetry energy $\epsilon_{sym}$ given in ref.\cite{Br68} yields $S(\rho) = 0.5 \epsilon_{sym}$ = 28 MeV and $S_1(\rho) = -S(\rho) \lambda$ = -18.76 MeV which are independent of nuceonic density.
 
\section{The liquid droplet model of nuclei and symmetry energy}
\label{section3}

      The volume and surface terms in the standard semi empirical mass formula pertain to the isospin symmetric systems. The volume coefficient provides the binding energy per nucleon whereas the surface coefficient, up to a certain extent, provides the surface energy. The symmetry term in the standard binding energy formula has a volume character only. But when the surface energy is affected by the isospin asymmetry, the thermodynamic consistency requires that some of the asymmetry moves to the surface. Minimization of the net nuclear energy with respect to the partitioning of asymmetry produces an expression \cite{Da03} for the binding energy B(A,Z) of a nucleus with mass number A and the atomic number Z given by  

\begin{equation}
 B(A,Z) = a_vA-a_sA^{2/3}-a_c\frac{Z(Z-1)}{A^{1/3}}-\frac{S_v}{1+\frac{S_v}{S_s} A^{-1/3}}\frac{(N-Z)^2}{A}+\delta,
\label{seqn8}
\end{equation}
\noindent 
where $\delta=a_pA^{-1/2}$ for even N-even Z, $-a_pA^{-1/2}$ for odd N-odd Z, 0 for odd A, and neutron number N=A-Z. The above expression is similar to the droplet model where skin size is a basic parameter and one of the starting points. $S_v$ and $S_s$ are now the volume and surface symmetry parameters, respectively, whereas $a_v$, $a_s$, $a_c$ and $a_p$ are the usual volume, surface, coulomb and pairing energy coefficients. Allowing the mass number A going to infinity, it may be seen that the volume symmetry energy coefficient $S_v$ is equal to the NSE obtained from the (infinite) nuclear matter calculation. Therefore extracting $S_v$ from measured atomic mass excesses provide experimental value for the NSE at normal nuclear density. Theoretical atomic mass excesses $\Delta M_{A,Z}$ can be obtained from the theoretical binding energy B(A,Z) by correcting for the electronic binding energy as

\begin{equation}
 \Delta M_{A,Z} = Z \Delta m_H  + (A-Z) \Delta m_n - a_{el} Z^{2.39} - b_{el} Z^{5.35} - B(A,Z)
\label{seqn9}
\end{equation}
\noindent
where $\Delta m_H = m_p + m_e - u$ =  7.28897050 MeV + $a_{el}$ + $b_{el}$ and $\Delta m_n = m_n - u$ = 8.07131710 MeV, $m_p$, $m_n$, $m_e$ are the masses of proton, neutron and electron and $u$ is the atomic mass unit, all expressed in MeV and the electronic binding energy constants \cite{Lu03} $a_{el} = 1.44381 \times 10^{-5}$ MeV and $b_{el} = 1.55468 \times 10^{-12}$ MeV. This approach \cite{Da03} also yields a relationship among neutron skin, $a_c$, $S_v$ and $S_s$ given by  

\begin{equation}
 \frac{R_n-R_p}{R} = \frac{A}{6NZ} \frac{N-Z-\frac{a_c}{12S_v}ZA^{2/3}}{1+\frac{S_s}{S_v} A^{1/3}}
\label{seqn10}
\end{equation}
\noindent    
The difference between equivalent sharp radii for neutrons $R_n$ and protons $R_p$ is primarily linear in the asymmetry and the symmetry coefficient ratio $S_v/S_s$ measures the neutron skin of a nucleus.

\section{Results and discussion}
\label{section4}

      We extract the nuclear symmetry energy $S_v$ from measured atomic mass excesses and associated errors using the maximum likelihood method described in detail in ref.\cite{Mo95}. This leads to the generalised equations\cite{Mo95},

\begin{equation}
\sum_{i=1}^n\frac{ [ \Delta M^i_{ex} - (\Delta M^i_{th}+{\mu_{th}}^*) ]}{{\sigma^{i}_{ex}}^2 + {\sigma_{th}}^{2*}}\frac{\partial{(\Delta M^i_{th})}}{\partial{p_{\nu}}} = 0, ~~~~~\nu=1,2,.......m 
\label{seqn11}
\end{equation}

\begin{equation}
\sum_{i=1}^n\frac{ [ \Delta M^i_{ex} - (\Delta M^i_{th}+{\mu_{th}}^*) ]^2-({\sigma^{i}_{ex}}^2 + {\sigma_{th}}^{2*})}{({\sigma^{i}_{ex}}^2 + {\sigma_{th}}^{2*})^2} = 0
\label{seqn12}
\end{equation}

\begin{equation}
\sum_{i=1}^n\frac{ [ \Delta M^i_{ex} - (\Delta M^i_{th}+{\mu_{th}}^*) ]}{({\sigma^{i}_{ex}}^2 + {\sigma_{th}}^{2*})} = 0, 
\label{seqn13}
\end{equation}  
\noindent
where $p_{\nu}$ are the unknown $m$ parameters of the model. Here $\Delta M^i_{ex}$ is the measured mass excess for a particular nucleus for proton number $Z$ and neutron number $N$, and $\Delta M^i_{th}$ is the corresponding calculated quantity and $\sigma^i_{ex}$ is the associated error in each of $n$ such measurements. $\sigma_{th}$ is the intrinsic model error which accounts for known and unknown missing terms in the theoretical model used for fitting the mass excesses. Here we assume that  the true mass excess $u_{tr}^i$ of the nucleus $i$ can be written as $u_{tr}^i=\Delta M^i_{th}+e_{th}^i$, where $e_{th}^i$ is the theoretical error term and is distributed normally as $e_{th}^i \in N(\mu_{th},\sigma_{th})$ with a mean $\mu_{th}$ and a standard deviation $\sigma_{th}$ around this mean. The notations ${\sigma_{th}}^{2*}$ and ${\mu_{th}}^*$  mean that by solving Eqs.(12,13) we obtain the estimates ${\sigma_{th}}^{2*}$ and ${\mu_{th}}^*$ of the true ${\sigma_{th}}^{2}$and $\mu_{th}$. Use of root-mean-square deviation ($\sigma_{rms}$) defined as

\begin{equation}
\sigma_{rms}=[\frac{1}{n}\sum_{i=1}^n ( \Delta M^i_{ex} - \Delta M^i_{th} )^2]^{1/2}
\label{seqn14}
\end{equation}
\noindent	
as the error of the theoretical mass model and obtained by minimising its value by adjusting the model parameters is reasonable when all the errors $\sigma^i_{ex}$ associated with the measurements are small compared to the model error $\sigma_{rms}$. However, for large experimental errors $\sigma^i_{ex}$ this definition is unsatisfactory, since both the theoretical and the experimental errors contribute to the rms deviation. We must therefore use an approach that decouples the theoretical and the experimental errors. In the present case, since it is a minimally modified formula without the shell corrections or the Wigner term, the theoretical model error must account for the various other known and unknown terms in the model. The model error ($\sigma_{th}$) obtained in this way (ML estimation) contains no contributions from the experimental uncertainties $\sigma^i_{ex}$. Thus,  we have two additional equations here compared to usual least square equations (minimising $\sigma^2$ or alternatively $\chi^2$)  that arise when model parameters are estimated by adjustments to experimental data under the assumption of a perfect theory with $\sigma_{th}=0$ and $\mu_{th}=0$. The above equations  are equivalent to minimizing $S$ with respect to $p_{\nu}$, where

\begin{equation}
S=\sum_{i=1}^n\frac{[\Delta M^i_{ex} - (\Delta M^i_{th}+{\mu_{th}}^* )]^2}{{\sigma^{i}_{ex}}^2 + {\sigma_{th}}^{2*}}
\label{seqn15}
\end{equation}
\noindent
and solving 

\begin{equation}
{\sigma_{th}}^{2*}=\frac{1}{\sum_{i=1}^n w_i^{k_{\sigma}}}
\sum_{i=1}^n w_i^{k_{\sigma}}
[{(\Delta M^i_{ex} - \Delta M^i_{th}-{\mu_{th}}^*)}^2 -{\sigma^{i}_{ex}}^2]
\label{seqn16}
\end{equation}

\begin{equation}
{\mu_{th}}^{*}=\frac{1}{\sum_{i=1}^n w_i^{k_{\mu}}}
\sum_{i=1}^n w_i^{k_{\mu}}
[(\Delta M^i_{ex} - \Delta M^i_{th})]	
\label{seqn17}
\end{equation}
\noindent
where
\begin{eqnarray}
w_i^k &=& \frac{1}{({\sigma^{i}_{ex}}^2 + {\sigma_{th}}^{2*})^k}  \\
k_{\sigma}&=& 2 \\
k_{\mu} &=& 1
\end{eqnarray}
\noindent
The unknowns ${\mu_{th}}^*$ and ${\sigma_{th}}^{2*}$ are then determined from Eqs.(16,17) by an iterative procedure whose convergence is found to be quite good. This way the experimental error is subtracted from the difference between the experimental and the calculated mass excesses. As the model considered here does not contain any term like $a_0A^0$, that is strictly a constant parameter, the most complete characterisation of the theoretical error requires both its mean $\mu_{th}$ and its standard deviation $\sigma_{th}$ around this mean. Hence we need to solve the full $m+2$ set of equations. If ${\mu_{th}}^*$ is found to be significantly different from zero the theory will need modification.  

      In ref.\cite{Ba04} both the chi-square and the sum of deviation squares were minimized. The results of those two minimization did differ slightly. However, the data are not expected to approach the theory when measurement errors tend to zero. In the present work the theoretical errors are assumed to accompany the experimental errors. In fitting experimental data, the theoretical errors are estimated simultaneously with the optimal parameter values. This procedure effectively produces a minimization which is intermediate between the minimization of the sum of deviation squares and that of the chi-square. This procedure gives the possibility for a realistic estimation of the uncertainties in the fitted parameters. As one may compare with ref.\cite{Ba04} to find that the volume symmetry energy coefficient obtained in the present work is substantially different. This is due to the fact that Bethe-Weizs\"acker mass formula is minimally modified along the lines of the liquid droplet model by partitioning the symmetry term into volume and surface terms. Other energy coefficients do not differ much beyond the limits of the earlier work \cite{Ba04}. 

      The coefficients of the liquid droplet model [Eq.(8)] are evaluated by fitting the recent measured and extrapolated atomic mass excesses from Audi-Wapstra-Thibault atomic mass table \cite{Au03} by minimizing $S$ of Eq.(15) and are shown in Table-1. The $\sigma_{th}$ and $\mu_{th}$ simultaneously solved from Eqs.(16,17) are also tabulated. The values of $\sigma_{th}$ and $\mu_{th}$ for 2228 experimentally measured atomic mass excesses are 2.880  and 0.029 respectively. Exclusion of the measured atomic mass excesses of lighter nuclei having mass number A $<$ 16 results in $\sigma_{th}$ = 2.782 and $\mu_{th}$ = 0.037. When the additional 951 extrapolated data are included for the same analysis, that is for total 3179 measured+extrapolated data, the values obtained are $\sigma_{th}=2.960$ and $\mu_{th}=0.040$. These values are acceptable as the values of $\mu_{th}$ do not differ significantly from zero. Fig.1 and Fig.2 show the plots of fitting errors of atomic mass excesses versus mass numbers for the droplet model [Eqs.(8,9)] mass formula. 

\begin{table}
\caption{Coefficients of the liquid droplet model mass formula extracted from atomic mass excesses.}
\centering
\begin{tabular}{|c|c|c|c|c|c|} 
\hline
\hline 
$a_v$ & $a_s$ & $a_c$ & $S_v$ &$S_s$& $a_p$  \\
MeV & MeV & MeV & MeV & MeV &MeV \\ \hline
&&&&&\\
$[a]~$15.563&17.652&0.695&29.687&17.680&11.813\\ 
$\pm$0.00043&$\pm$0.00107&$\pm$0.00004&$\pm$0.0097&$\pm$0.0216&$\pm$0.0044\\ \hline
&&&&&\\
$[b]~$15.500&17.480&0.689&30.048&16.674&10.246\\ 
$\pm$0.00012&$\pm$0.00025&$\pm$0.00002&$\pm$0.0042&$\pm$0.0071&$\pm$0.00048\\ \hline
&&&&&\\
$[c]~$15.465&17.394&0.686&30.130&16.317&10.273 \\ 
$\pm$0.00012&$\pm$0.00025&$\pm$0.00002&$\pm$0.0043&$\pm$0.0067&$\pm$0.00049\\ \hline
\hline 
\end{tabular} 
\vskip 0.2cm
$[a]$. Using experimentally measured 2168 atomic mass excesses for A$\ge$16. 
$\mu_{th}=0.037$ and $\sigma_{th}=2.782$\\         
$[b]$. Using all the experimentally measured 2228 atomic mass excesses. 
$\mu_{th}=0.029$ and $\sigma_{th}=2.880$\\         
$[c]$. Using measured 2228 + extrapolated 951 atomic mass excesses. 
$\mu_{th}=0.040$ and $\sigma_{th}=2.960$\\

\end{table}
\noindent

\begin{figure}[htbp]
\eject\centerline{\epsfig{file=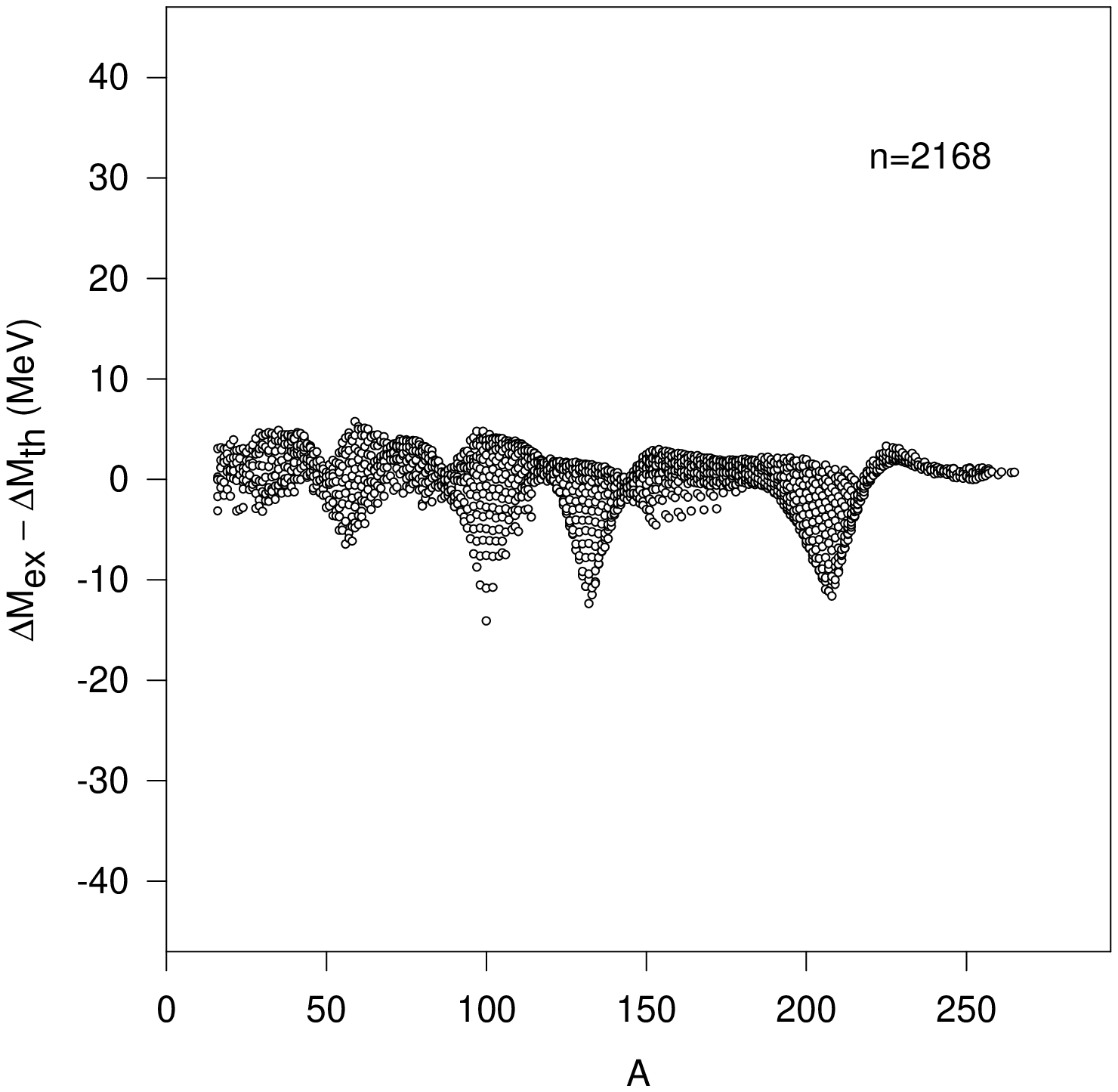,height=10cm,width=10cm}}
\caption
{The plot of differences between 2168 (A$\ge$16) measured and theoretical atomic mass excesses calculated by the liquid droplet model [Eq.(8)] mass formula versus mass number A.}
\label{fig1}
\end{figure}

\begin{figure}[htbp]
\eject\centerline{\epsfig{file=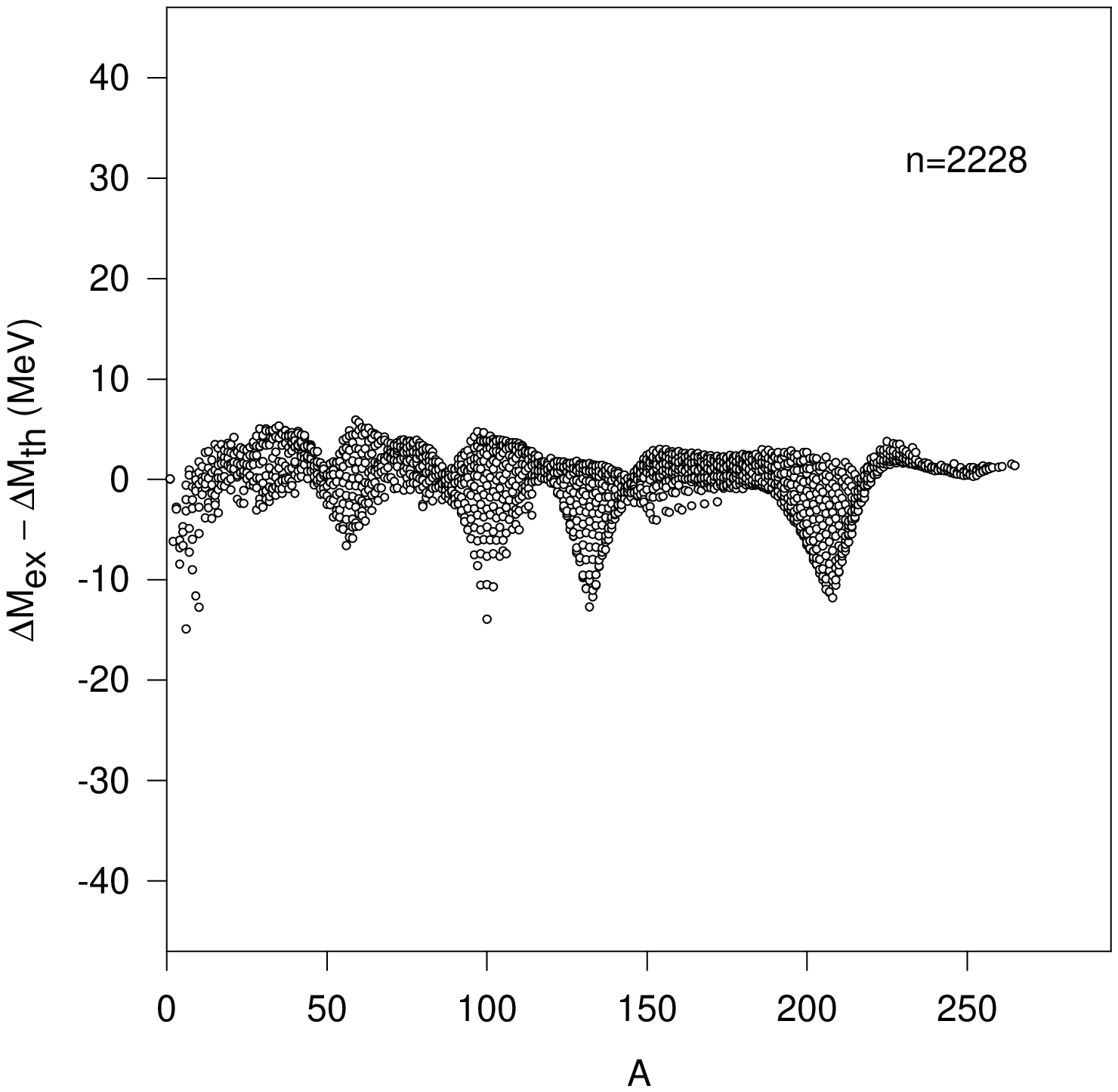,height=10cm,width=10cm}}
\caption
{The plot of differences between measured 2228 and theoretical atomic mass excesses calculated by the liquid droplet model [Eq.(8)] mass formula versus mass number A.}
\label{fig2}
\end{figure}
\vspace{-0.14cm}

      A mean field calculations with DDM3Y effective interaction is performed using the usual values of energy dependence parameter $\alpha= 0.005 MeV^{-1}$ \cite{Sa79}, the saturation density $\rho_0=0.1533 fm^{-3}$ \cite{BSD90} and the saturation energy per nucleon $\epsilon_0=-15.26 \pm 0.52$ MeV. This saturation energy per nucleon is the volume energy coefficient and this value used in the present calculations more or less covers the entire range of values of $a_v$ obtained here (Table-1) or fitting the original Bethe-Weizs\"acker mass formula \cite{Ba04} or other droplet models \cite{Mo95,St05,My74,Li82,Kr83}. The value obtained for the NSE at the saturation density, $S(\rho_0)$, is found to be $31.18 \pm 0.29$ MeV using definition of Eq.(1) and $31.86 \pm 0.29$ MeV using definition of Eq.(7). 
       
      The value of $S_v=30.048 \pm 0.004$ MeV extracted from experimental mass excesses is reasonably close to the theoretical estimate of the value of NSE at the saturation density $S(\rho_0)$ described above. For A $\ge$ 16 little change is observed to the value of $S_v$ which becomes $29.687 \pm 0.010$ whereas the $S_s$ changes to a larger extent to $17.680 \pm 0.022$. This behaviour suggests the fact that the surface energy depends upon symmetry. The value obtained for $S_v$ in ref. \cite{Da03} is between 29.10 MeV to 32.67 MeV and that obtained by the liquid droplet model calculation of ref. \cite{St05} is 27.3 MeV whereas in ref. \cite{Di05} it is 28.0 MeV. It should be mentioned that the value of the volume symmetry energy co-efficient $S_v$ in some advanced mass description \cite{Po03} is close to the present value which with their $-\kappa_{vol}.b_{vol}=S_v$ equals 29.3 MeV. In a very recent similar work \cite{Ro06} the volume symmetry energy co-efficient comes out to be about 29 MeV when with other terms, pairing energy, Wigner term and shell corrections are also included. The ratio $S_v /S_s$ which is a measure of the neutron skin thickness is found to be about 1.8 in the present calculations. The value of this ratio obtained in ref. \cite{Da03} is about 2.0 to 2.8 whereas in liquid droplet model calculations of ref. \cite{St05} the value obtained is 1.68 and in ref. \cite{Di05} it is calculated to be 1.3.    

\begin{table}[h]
\caption{Result for the $S(\rho_0)$ of the present mean field calculation is compared with the results from the variational calculations of Refs. \cite{Ak98,Wi88} using the Argonne and Urbana NN potentials, in combination with Urbana models for the {\it TNI}. The last column includes a relativistic boost correction $\delta v$ and the adjusted UIX$^*$ {\it TNI}.}
\centering
\begin{tabular}{|c|c|c|c|} 
\hline 
\hline 
Present Calc.&Av14&Av14+UVIII&Uv14 \\ 
MeV & MeV & MeV & MeV \\ \hline
[a]~$31.18 \pm 0.29$&24.90&27.49&26.39 \\ \hline
[b]~$31.86 \pm 0.29$\\ \hline
\hline 
Uv14+UVIII&Av18&Av18+UIX&Av18+$\delta v$+UIX$^*$ \\ 
MeV &MeV &MeV &MeV \\ \hline
28.76&26.92&29.23&30.1 \\ \hline
\hline 
\end{tabular} 
\vskip 0.2cm
$[a]$. Using definition of Eq.(1) for the nuclear symmetry energy. \\       
$[b]$. Using definition of Eq.(7) for the nuclear symmetry energy. 
\end{table}

      The value of NSE at nuclear saturation density $\approx$ 30 MeV, therefore, seems well established empirically. However, theoretically different sets of parametrizations of the relativistic mean-field (RMF) models, which fit the obseravables for isospin symmetric nuclei well, lead to a relatively wide range of predictions 24-40 MeV for $S(\rho_0)$. In Table-2 results for the $S(\rho_0)$ using DDM3Y interaction are compared with the results from the variational calculations using the Argonne and Urbana NN potentials, in combination with Urbana models for the three-nucleon interaction [{\it TNI}]. The last column includes a relativistic boost correction and the adjusted UIX$^*$ {\it TNI}. The present result of the mean field calculation is close to the result of Av18+$\delta v$+UIX$^*$ variational calculation \cite{Ak98}.

\section{Summary and conclusion}
\label{section5}

      In summary, we show that theoretical description of nuclear matter based on mean field calculation using density dependent M3Y effective NN interaction gives a value of the symmetry energy that is consistent with the empirical value extracted by fitting the droplet model to the measured and extrapolated atomic mass excesses using the maximum likelihood estimator method. The value of volume symmetry energy coefficiet changes little compared to the surface symmetry energy coefficient when measured atomic mass excesses of nuclei lighter than 16 are excluded. This observation highlights the fact that the symmetry energy depends on the surface tension and {\it vice versa}. The volume and surface symmetry energy coefficients are related to the neutron skin of finite nucleus. Such mean field calculations of nuclear symmetry energy thus satisfy the constraints from finite nuclei and also agree with recent theoretical descriptions of nuclear matter.

\end{document}